\documentclass[conference, anonymous]{IEEEtran}
\IEEEoverridecommandlockouts

\usepackage{cite}
\usepackage{amsmath,amssymb,amsfonts}
\usepackage{textcomp}
\usepackage{xspace}
\usepackage{xcolor,colortbl}
\usepackage[most]{tcolorbox}
\usepackage{tabularx}
\usepackage{newfloat}
\usepackage{listings}
\usepackage{soul}
\usepackage{color}
\usepackage{caption} 
\usepackage{multirow}
\usepackage{graphicx}
\usepackage{graphicx}
\usepackage{subcaption}
\usepackage{algpseudocode}
\usepackage[linesnumbered,ruled,vlined]{algorithm2e}

\SetKwInput{KwInput}{Input}
\SetKwInput{KwOutput}{Output}

\def\BibTeX{{\rm B\kern-.05em{\sc i\kern-.025em b}\kern-.08em
    T\kern-.1667em\lower.7ex\hbox{E}\kern-.125emX}}

\newtcolorbox{boxEnv}{
    center,
    left= 0.5 mm,
    top = 0.25 mm,
    right = 0.5 mm,
    bottom =0.25 mm,
    colframe=gray!90!black,
    colback=black!5!white, 
    boxrule=0.5pt,
    title = Anchor Generation Prompt:
}

\begin{document}

\title{Semantic-Aware Contrastive Fine-Tuning: Boosting Multimodal Malware Classification with Discriminative Embeddings\\}


\author{\IEEEauthorblockN{Ivan Montoya Sanchez\IEEEauthorrefmark{1}, Shaswata Mitra\IEEEauthorrefmark{2}, 
Aritran Piplai\IEEEauthorrefmark{3},
Sudip Mittal\IEEEauthorrefmark{4}
}
\IEEEauthorrefmark{1}\IEEEauthorrefmark{3}The University of Texas at El Paso, TX, USA\\
\IEEEauthorrefmark{2}\IEEEauthorrefmark{4}Mississippi State University, MS, USA
\\ \IEEEauthorrefmark{1}iamontoyasa@miners.utep.edu, \IEEEauthorrefmark{2}sm3843@msstate.edu, \IEEEauthorrefmark{3}apiplai@utep.edu,
\IEEEauthorrefmark{4}mittal@cse.msstate.edu\\
}

\maketitle

\begin{abstract}
    The rapid evolution of malware variants requires robust classification methods to enhance cybersecurity. While Large Language Models (LLMs) offer potential for generating malware descriptions to aid family classification, their utility is limited by semantic embedding overlaps and misalignment with binary behavioral features. We propose a contrastive fine-tuning (CFT) method that refines LLM embeddings via targeted selection of hard negative samples based on cosine similarity, enabling LLMs to distinguish between closely related malware families. Our approach combines high-similarity negatives to enhance discriminative power and mid-tier negatives to increase embedding diversity, optimizing both precision and generalization. Evaluated on the CIC-AndMal-2020 and BODMAS datasets, our refined embeddings are integrated into a multimodal classifier within a Model-Agnostic Meta-Learning (MAML) framework on a few-shot setting. Experiments demonstrate significant improvements: our method achieves 63.15\% classification accuracy with as few as 20 samples on CIC-AndMal-2020, outperforming baselines by 11--21 percentage points and surpassing prior negative sampling strategies. Ablation studies confirm the superiority of similarity-based selection over random sampling, with gains of 10-23\%. Additionally, fine-tuned LLMs generate attribute-aware descriptions that generalize to unseen variants, bridging textual and binary feature gaps. This work advances malware classification by enabling nuanced semantic distinctions and provides a scalable framework for adapting LLMs to cybersecurity challenges.
\end{abstract}


\section{Introduction}

    Rapidly evolving malware threats present a significant challenge in critical infrastructure, with numerous new variants emerging daily. According to Statista ~\cite{ststista_malware_variants}, approximately 465,500 malware variants were reported in 2022 alone. These variants share common characteristics, codes, or behavioral patterns and are classified into malware families. Each variant represents a different version or modification in a specific malware family, featuring slight alterations in their code or behavior designed to evade detection or enhance effectiveness. Classifying malware families helps security systems detect and respond to new variants based on the known characteristics of each family to develop or update security measures. Thus, categorizing these unrecognized variants of malware is essential for effective cybersecurity.

    LLMs offer a promising solution by generating textual descriptions of new malware variants, which can be valuable for classifying malware families. However, current LLMs often struggle to align with structured binary features, limiting the effectiveness of these generated descriptions. CFT addresses this challenge by optimizing embedding spaces—bringing similar samples or descriptions closer together and pushing dissimilar ones apart. To be effective, CFT requires more sophisticated methods for selecting dissimilar or negative samples than traditional heuristic approaches, such as random sampling or simple category-based techniques. These conventional methods are often inadequate for distinguishing closely related and disparate malware families, especially when their semantic embeddings overlap significantly. To clarify our research problem, we consider providing the following use-case for better reader understanding. 
    
    Consider a critical infrastructure seeking to defend against malware threats but lacking access to all relevant malware samples—since many organizations do not disclose their cyber incidents. However, textual cyber threat intelligence (CTI) reports are widely shared across sources. LLMs, as few-shot learners \cite{cahyawijaya2024llms}, can process these descriptions and transform them into structured representations usable by downstream cyber-defense models, enhancing malware family identification through embedding similarity. This classification method via embedding similarity is well established in other domains, such as computer vision, where textual descriptions from language models significantly enhance object identification ~\cite{radford2021learning}. In the embedding space of textual descriptions, it is easy to distinguish between different object concepts. For example, if an image description includes the word \textit{``dog"} or a particular \textit{``dog breed"}, the embeddings will vary significantly compared to a description that includes \textit{``wolves"}. However, in cybersecurity, the concepts of different types of malware often overlap, making classification less effective. As a result, classifying malware based on behavior descriptions presents challenges for downstream tasks.

    To overcome this issue, we introduce an improved CFT method for selecting more informative and challenging hard negatives specifically tailored for malware description generation. Our approach involves using cosine similarity between embeddings as the primary criterion for negative selection, intentionally choosing negatives that closely match the semantic similarity observed between positive samples across different malware families. By selecting negatives within a carefully chosen high-similarity range, we ensure the model learns to make finer semantic distinctions, resulting in embeddings that better differentiate malware families. In addition, we aim to fine-tune the model so it can generalize effectively to new, previously unseen malware samples. Specifically, our approach encourages the model to generate descriptions that combine a general understanding of the malware family with the specific behavioral attributes observed in an unseen instance. This capability significantly enhances the practical utility of LLM-generated descriptions by ensuring that the model produces semantically accurate family-level descriptions and aligns closely with the unique features of each new malware sample it encounters.

    \begin{figure}[t]
        \centering
        \includegraphics[width=\linewidth]{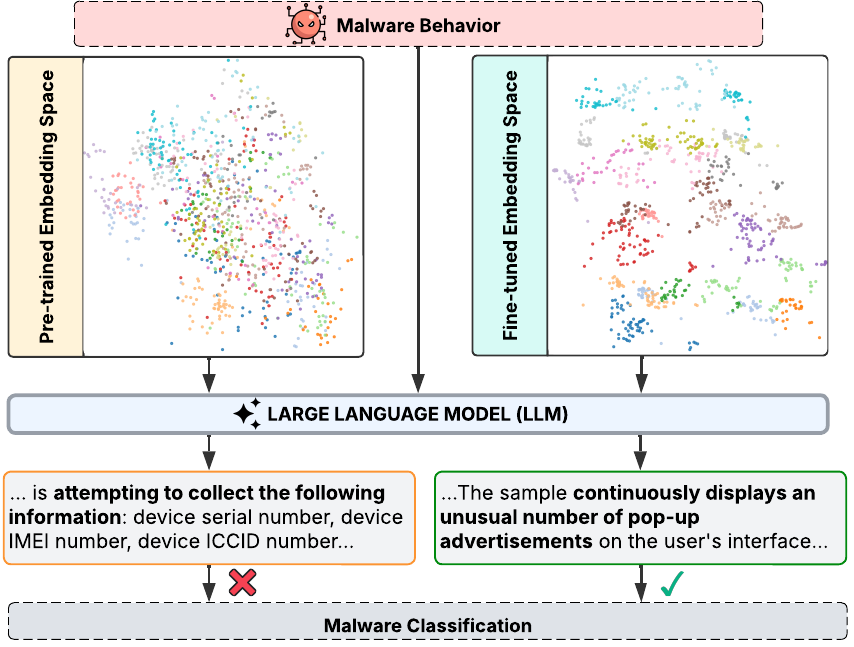}
        \caption{\small Overview of our similarity-based contrastive fine-tuning framework for malware classification. Initially, embeddings from a pre-trained LLM exhibit significant overlap among malware families, leading to ambiguous descriptions and poor classification. In contrast, in similarity-based contrastive fine-tuning, embeddings become discriminative, clearly separating malware families into distinct clusters. This improved embedding space enables the LLM to generate precise, attribute-specific malware descriptions, substantially enhancing malware classification accuracy.}
        \label{fig:tldr}
    \end{figure}
    
    To empirically validate our approach, we perform experiments using two widely recognized malware datasets, CIC-AndMal-2020~\cite{cicandmal2020} and BODMAS~\cite{bodmas}, which contain diverse malware families and challenging classification scenarios. To demonstrate the practical effectiveness of our embeddings, we integrate them into a multimodal malware classifier within a Model-Agnostic Meta-Learning (MAML) framework ~\cite{finn2017model}. Combining embeddings generated from our improved CFT method with dynamic binary attributes, we demonstrate clear performance improvements over existing baseline models relying solely on binary features. To the best of our knowledge, we made the first attempt at malware classification using multimodal techniques, where one mode involved behavior features and the other included textual descriptions.

    Figure~\ref{fig:tldr} illustrates the primary motivation and impact of our contrastive fine-tuning approach. Initially, the Pre-trained embedding space generated by an LLM exhibits significant overlap among malware families, limiting the discriminative quality of generated descriptions. In contrast, our similarity-based contrastive fine-tuning method produces a refined embedding space, clearly separating malware families into distinct semantic clusters. Consequently, the fine-tuned LLM generates precise and discriminative malware descriptions, improving malware classification performance.

    Our contributions are the following:

    
    \begin{itemize} 
        \item We developed a novel algorithm to select more challenging hard negatives in CFT while significantly reducing the semantic overlap between dissimilar malware variant embeddings. 
        \item We demonstrate that post-fine-tuning, LLMs can generate accurate, attribute-aware malware descriptions that generalize well to unseen samples. 
        \item We empirically validate improved embedding quality in a few-shot setting through downstream classification accuracy, significantly surpassing existing methods for negative selection.
    \end{itemize}

    In Section \ref{section:related_work}, we discuss related works and provide necessary background information. In Section \ref{section:methodology}, we offer a detailed description of our research approach. The experiments and evaluations are presented in Section \ref{section:results}. Finally, Section \ref{section:conclusion} includes concluding remarks and directions for future research.

    
\section{Related Work}\label{section:related_work}

    \subsection{Contrastive learning} 
        Contrastive learning methods learn effective embeddings by pulling similar samples closer together and pushing dissimilar ones apart. Recent approaches such as Supervised Contrastive Learning for Pre-trained Language Model Fine-Tuning~\cite{gunel2020supervised} and CLIP~\cite{radford2021learning} demonstrated the effectiveness of contrastive learning across vision and language domains. However, these methods for multimodal alignment often fail to address the nuanced overlaps inherent to malware classification, where families frequently share code or behaviors.

        The widely used InfoNCE loss~\cite{oord2018representation} remains central to many contrastive approaches and depends heavily on negative sample selection. Random negative sampling remains common due to simplicity, but it often leads to suboptimal performance on highly specific tasks. Improved heuristic methods include selecting negatives based on class labels~\cite{khosla2020supervised} or semantic clustering~\cite{robinson2020contrastive}. However, these heuristics are inadequate for tasks with high semantic overlap among classes, such as malware family classification.

Recent work by Xu et al.~\cite{xu2024enhancing} introduces a distance-aware approach to contrastive learning, where the definition of positive and negative samples is softened based on their relative distances in the embedding space. This method leverages a weighted feature-distance calculation, effectively creating a continuous spectrum between positive and negative examples instead of a rigid binary separation. Although effective, their approach primarily addresses cross-domain few-shot learning in visual tasks without accessing labeled data, whereas our methodology focuses explicitly on fine-grained semantic distinctions critical for malware family classification, utilizing cosine similarity to systematically select the most challenging negative examples.

In contrast, our method explicitly selects negatives based on high cosine similarity with embeddings generated by pre-trained LLMs, ensuring negatives are particularly challenging and informative. This tailored negative sampling enhances the ability of models to discern subtle semantic differences among closely related malware families, directly addressing the semantic overlap challenge unique to cybersecurity contexts.
    \subsection{Malware description generation} 
        Malware description generation has recently leveraged LLMs such as GPT and LLaMA~\cite{touvron2023llama} to generate narratives describing malware behaviors. Prior work~\cite{alkhatib2023gpt} shows these descriptions improve analyst productivity, but they often lack alignment with structured binary features (e.g., API calls, registry modifications), limiting their utility in automated classification. For instance, generic LLM outputs may fail to emphasize family-specific traits like encryption routines or persistence mechanisms. This misalignment causes embedding overlaps, where distinct families appear semantically similar in LLM-generated representations. Our contrastive fine-tuning framework directly addresses this by refining embeddings to accentuate discriminative attributes, ensuring generated descriptions are both interpretable and machine-actionable.
    \subsection{Multimodal classification} 
        Multimodal classification combines diverse data types (e.g., text, binaries, behavioral logs) to improve malware detection. Early work by Shafiq et al.~\cite{shafiq2020multimodal} fused static and dynamic features, while Kim et al.~\cite{kim2020multimodal} integrated network traffic patterns with executable metadata. However, these efforts focus on low-level features, neglecting the semantic richness of textual descriptions. Recent studies~\cite{chen2022multimodal} explore LLM-derived text for malware analysis but face challenges in aligning free-form narratives with structured attributes. Our work bridges this gap by pairing contrastively fine-tuned embeddings with binary features in a unified latent space, enabling joint modeling of semantic and behavioral traits. Experiments demonstrate that this approach significantly improves classification accuracy over single-modality baselines, highlighting the value of aligning textual and structured modalities.experiments.
    \subsection{Meta-adaptation and Knowledge-Distillation} 
        Meta-learning, particularly Model-Agnostic Meta-Learning (MAML)~\cite{finn2017model}, enables rapid adaptation to new tasks with limited data—a critical capability for detecting novel malware variants. In cybersecurity, MAML has been applied to few-shot intrusion detection~\cite{yao2019meta} and dynamic malware analysis~\cite{wang2021meta}, primarily leveraging low-level behavioral features (e.g., API sequences). While effective, these prior frameworks lack mechanisms to integrate semantic context from textual descriptions, which can provide critical insights into malware intent and functionality. Our work introduces a novel fusion of MAML with contrastively optimized embeddings, enabling the classifier to rapidly adapt using both high-level semantic narratives and low-level behavioral attributes. This approach explicitly optimizes embeddings for meta-learning scenarios, ensuring that textual and binary modalities enhance generalization to unseen families, as demonstrated in our experiments.

        Meta-learning, particularly Model-Agnostic Meta-Learning (MAML)~\cite{finn2017model}, enables rapid adaptation to new tasks with limited data—a critical capability for detecting novel malware variants. In cybersecurity, MAML has been successfully applied to few-shot intrusion detection~\cite{yao2019meta} and dynamic malware analysis~\cite{wang2021meta}, primarily leveraging low-level behavioral features such as API sequences. While effective, these prior frameworks typically lack mechanisms to integrate semantic context from textual descriptions, which can provide critical insights into malware intent and functionality.

        Our work introduces a novel fusion of MAML with contrastively optimized embeddings, enabling rapid classifier adaptation that leverages both high-level semantic narratives and low-level behavioral attributes. To further enhance multimodal integration, we employ knowledge distillation~\cite{hinton2015distilling, gou2021knowledge}, a technique proven effective in transferring learned representations from a well-performing teacher model to a student model. Knowledge distillation is particularly advantageous over conventional feature-level fusion strategies—such as concatenation, attention-based fusion, or weighted averaging~\cite{baltruvsaitis2018multimodal,gao2020survey}—because it explicitly transfers predictive capabilities through soft labels. This results in more robust and interpretable fusion, particularly beneficial in scenarios with limited or noisy data~\cite{guo2022learning}. 

By leveraging a teacher model trained solely on behavioral attributes, our student model effectively combines structured binary features with semantically rich embeddings, achieving improved generalization and interpretability. This distillation-based approach thus provides a more effective and robust method of multimodal fusion compared to traditional feature integration techniques.

    \begin{figure*}[!t] 
        \centering
        \includegraphics[width=0.7\textwidth]{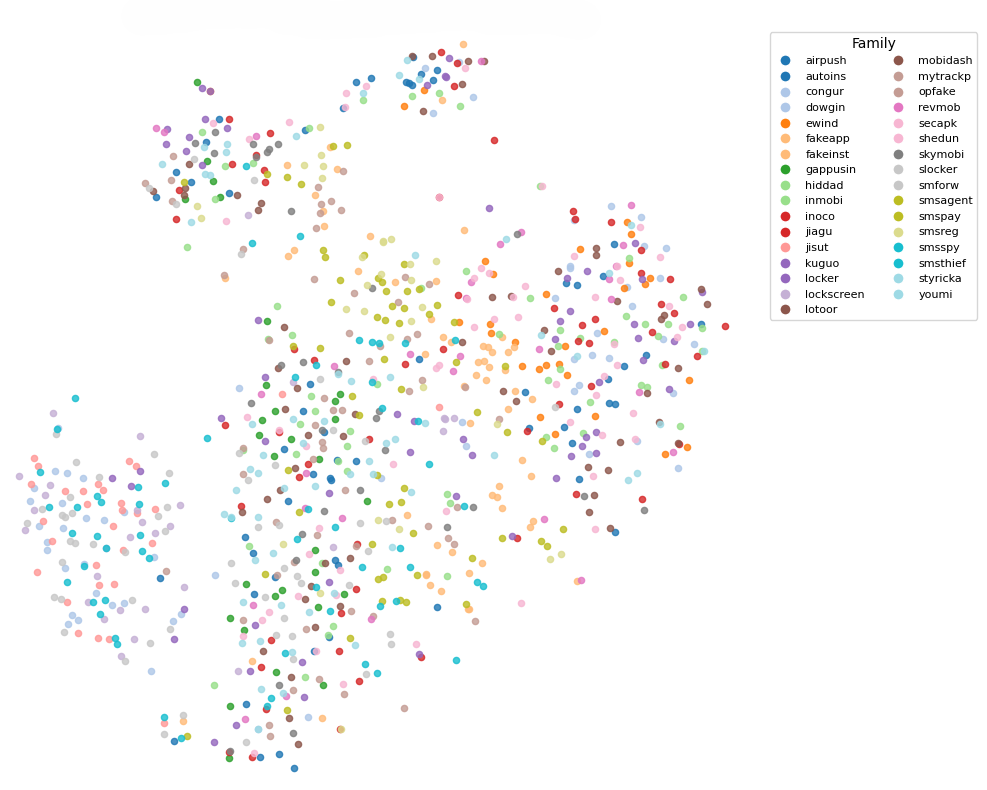} 
        \caption{\small Visualization of the pre-trained embedding space of malware descriptions generated by LLaMA-3.1-8B, projected into two dimensions using UMAP. Each color represents a different malware family. The significant overlap and lack of clearly defined clusters demonstrate the pre-trained model's limited capability to semantically distinguish among closely related malware families.} 
        \label{fig:example}
    \end{figure*}

\section{Methodology}\label{section:methodology}
    \subsection{Datasets}
        
        We conduct experiments using two malware datasets: CIC-AndMal-2020 ~\cite{cicandmal2020} and BODMAS ~\cite{bodmas}. The CIC-AndMal-2020 dataset comprises malware samples from multiple families, each described by 140 high-level behavioral features extracted via dynamic analysis. After removing families with insufficient samples-primarily from the zero-day category- we retain 33 malware families for evaluation.In contrast, the BODMAS dataset consists of 2,380 static binary features derived from low-level code characteristics. Following the feature reduction approach used in the EMBER dataset~\cite{2018arXiv180404637A}, we identify the most informative attributes using LightGBM's feature importance scores, which measure the contribution of each feature to the model's decision boundaries based on information gain and split frequency. We then select the top 64 features to reduce computational overhead during anchor and positive sample generation, while preserving the most discriminative information for downstream classification. To ensure consistency and robust evaluation, we also limit our analysis to a subset of 15 malware families.
    

    \subsection{Anchor and Positive Sample Generation}
        Anchors are generated by prompting four LLMs: LLaMA-3.2-1B\footnote{LLaMA-3.2-1B: huggingface.co/meta-llama/Llama-3.2-1B}, LLaMA-3.2-3B\footnote{LLaMA-3.2-3B: huggingface.co/meta-llama/LLaMA-3.2-3B}, LLaMA-3.1-8B\footnote{LLaMA-3.1-8B: huggingface.co/meta-llama/LLaMA-3.1-8B}, and Mistral-7B-v0.1\footnote{Mistral-7B-v0.1: huggingface.co/mistralai/Mistral-7B-v0.1}. Each model receives the same prompt containing the malware sample's binary or dynamic attributes.
    
        \begin{boxEnv}
            You are a cybersecurity expert specialized in malware detection. Imagine you received a malware sample with the following observations from behavioral analysis: \textcolor{teal}{\{attributes\}}. Describe these findings in terms of system attributes.
        \end{boxEnv}
    
    
        For the positive sample selection process, several general descriptions for the malware families are available. However, these descriptions need to be filtered because the model does not distinguish between malware families. Figure~\ref{fig:example} shows the . Although some of the families have visually defined clusters, more than one overlaps in the representation space, indicating that the embeddings alone fail to provide distinct semantic boundaries.. Such is the case for the \textbf{slocker} and \textbf{smsspy} families, or \textbf{smspay} and \textbf{smsreg} pair. Ideally, we want to filter the positive samples so that they do not overlap between families and are as close together as possible. To maintain high semantic coherence and minimize embedding overlap, we select exactly one ground-truth description per family. This choice ensures that intra-family cosine similarity remains consistently higher than inter-family similarities.
        We generate 200 model-inferred descriptions (anchors) per family for both datasets. Each anchor is paired with an expert-generated ground-truth description (positive sample) specific to its malware family.

        \begin{figure}[t]
            \centering
            \includegraphics[width=\linewidth]{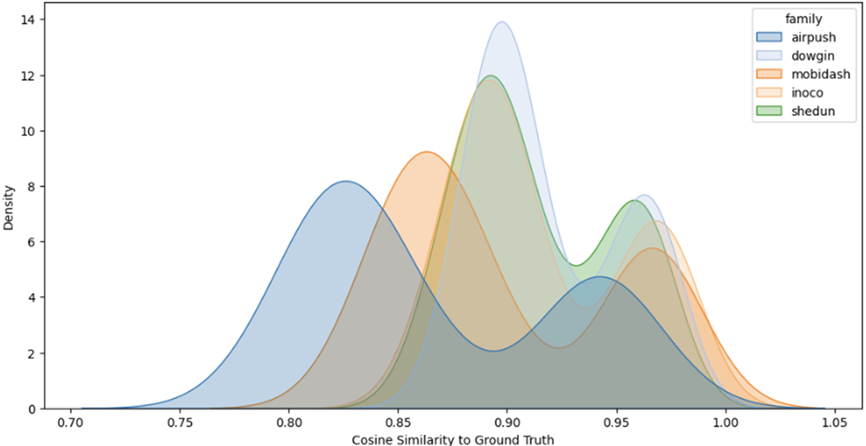}
            \caption{\small Distribution of cosine similarity scores between candidate negative samples and ground-truth descriptions for LLaMA-3.2-1B. Hard negatives are selected from the right end of the distribution, exhibiting the highest semantic similarity to the ground-truth. These samples create a more challenging CFT setting by forcing the model to distinguish between highly similar descriptions.}
            \label{fig:hard_neg_distribution}
        \end{figure}

    \subsection{Hard Negative Selection}
        
        We propose an improved hard-negative selection strategy to enhance model generalization. For each malware family, candidate negative descriptions from other families are first embedded using the pre-trained LLMs, deriving embeddings by mean pooling the final hidden-state representations. These embeddings are then ranked according to their cosine similarity with the family-specific ground-truth description embedding, also obtained from the pre-trained models. We select the top 20 negatives exhibiting the highest cosine similarity scores (typically ranging from 0.85 to 0.95), forming a set of challenging negatives. Figure~\ref{fig:hard_neg_distribution} shows the hard negative distribution for 5 families of the CIC-AndMal-2020 dataset. From this distribution we can detect if there are enough hard negatives with high cosine similarity and its quality. Additionally, we randomly select 12 mid-tier negative samples with lower similarity scores, ensuring broader semantic diversity.  
        This balanced approach, combining high-tier negatives to train the model in discriminating subtle semantic differences and mid-tier negatives to broaden coverage of the embedding space, promotes more robust and generalizable embeddings. Algorithm~\ref{alg:neg_mining} outlines the positive and negative sampling process.

        \begin{table}[!t]
            \caption{Description of Notations}
            \label{tab:maml_results_andmal}
            \centering
            \begin{tabular}{l|c}
                \hline
                \bfseries Notation & \bfseries Description \\ 
                \hline\hline
                $\mathcal{F} = \{f_1, \dots, f_N\}$ & Malware Family Set \\[1pt]
                $\mathcal{D}_{f_i}$ & Descriptions of samples from $f_i$\\[1pt]
                $d_i$ & One Positive Sample \\[1pt]
                $\mathcal{N}_{\text{hard}}$, $\mathcal{N}_{\text{diverse}}$ & Negative Pools \\[1pt]
                $ \emptyset $ & Null Set \\ [1pt]
                $T$ & Maximum Similarity Threshold\\ [1pt]
                \hline
            \end{tabular}
        \end{table}

        \begin{algorithm}[!t]
            \caption{ Positive and Negative Sample Selection }
            \label{alg:neg_mining}
            \KwInput{$\mathcal{D}_{f} \ \forall \ {f_i} | \ f_i \ \in \ \mathcal{F} $}
            \KwOutput{$d_i$, $\mathcal{N}_{\text{hard}}$, $\mathcal{N}_{\text{diverse}}$}
            
            \ForEach{$f_i \in \mathcal{F}$}{
                $\mathcal{N}_{\text{candidates}} \leftarrow \emptyset $ \\
                $d_i \leftarrow select\_sample(\mathcal{D}_{f_i}$)  \\
                $e_i \leftarrow \texttt{embedding}(d_i)$
                 \\
            
                \ForEach{$f_j \in \mathcal{F} \ \& \ i \neq j$}{
                    \ForEach{$d_j \in \mathcal{D}_{f_j}$}{
                        $e_j \leftarrow \texttt{embedding}(d_j)$ \\
                        $s \leftarrow cosine\_similarity (e_i, e_j)$ \\
                        \If{$s \leq T $}{
                            $\mathcal{N}_{\text{candidates}} \leftarrow  \ d_j \cup \ \mathcal{N}_{\text{candidates}}$
                        }
                    }
                }
            }
            $\mathcal{N}_{\text{candidates}} \leftarrow descending\_sort(\mathcal{N}_{\text{candidates}} \ , \ s$) \\
                $\mathcal{N}_{\text{hard}} \leftarrow get\_top\_20\_entries(\mathcal{N}_{\text{candidates}}$) \\
                $\mathcal{N}_{\text{diverse}} \leftarrow get\_any\_12\_samples(\mathcal{N}_{\text{candidates}} - \mathcal{N}_{\text{hard}} $) \\
                $\textbf{return} \ d_i, \ \mathcal{N}_{\text{hard}}, \ \mathcal{N}_{\text{diverse}} $
        \end{algorithm}

        During training sample generation, each anchor-positive pair is combined with 5 randomly chosen negatives from the high-tier set and 3 from the mid-tier set, resulting in four distinct contrastive training samples per anchor. Consequently, we generate 26,400 training samples for CIC-AndMal-2020 (33 families) and 12,000 samples for BODMAS (15 families).

    \begin{figure*}[!t]
      \centering
      \begin{subfigure}[b]{0.3\textwidth}
        \includegraphics[width=\textwidth]{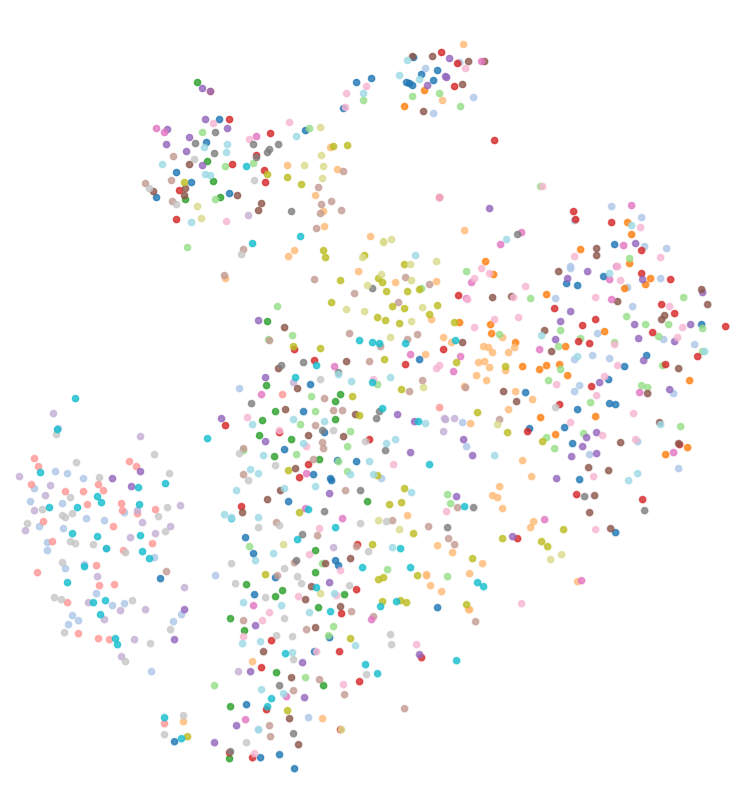}
        \caption{Pre-trained}
        \label{fig:pretrained}
      \end{subfigure}
      \hfill
      \begin{subfigure}[b]{0.3\textwidth}
        \includegraphics[width=\textwidth]{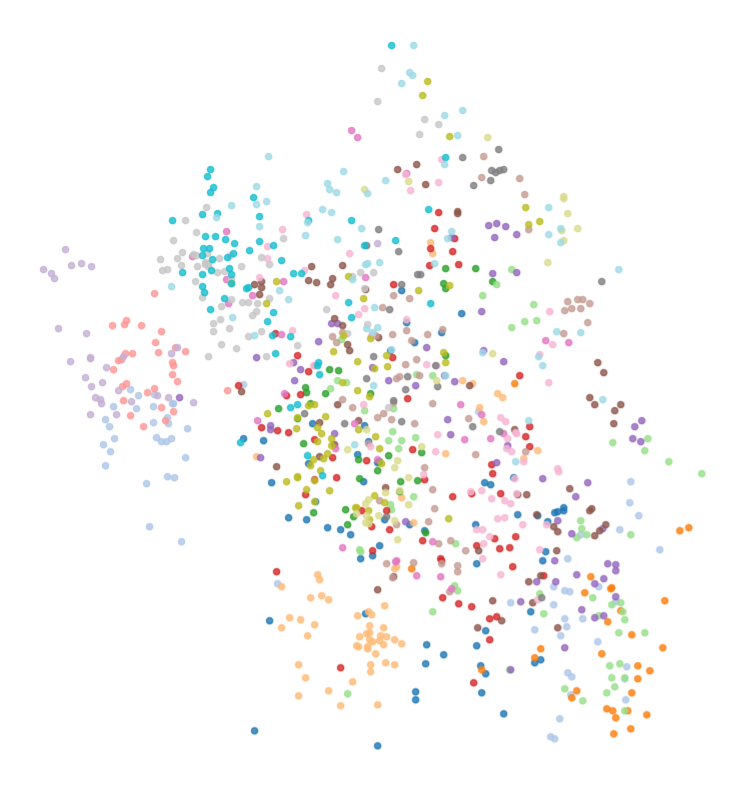}
        \caption{Random Negatives (CFT)}
        \label{fig:random_cft}
      \end{subfigure}
      \hfill
      \begin{subfigure}[b]{0.3\textwidth}
        \includegraphics[width=\textwidth]{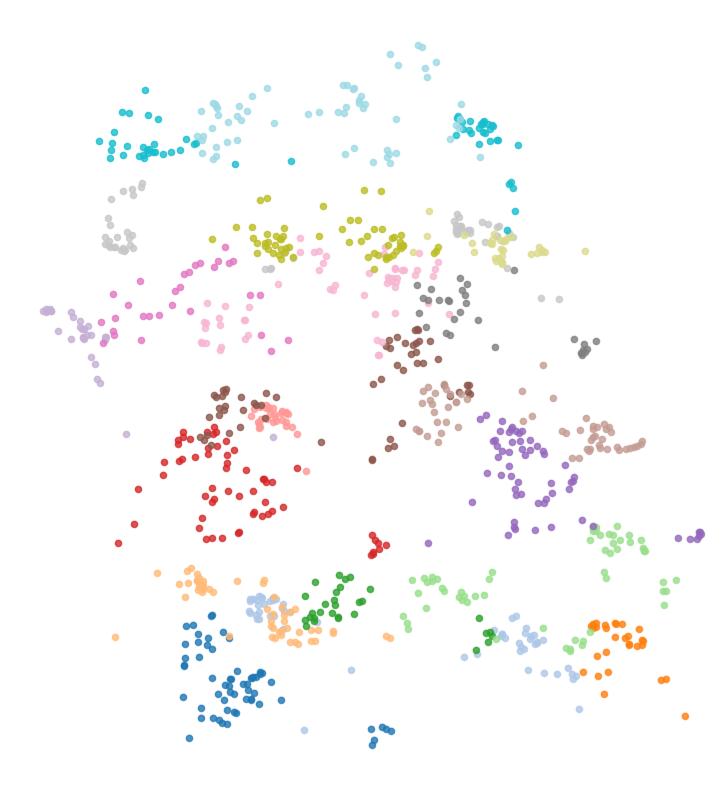}
        \caption{Similarity-based Negatives (CFT)}
        \label{fig:similarity_cft}
      \end{subfigure}
      \caption{\small Comparison of LLaMA-3.1-8B embedding spaces for CIC-AndMal-2020 dataset: pre-trained vs. contrastive fine-tuning with random and similarity-based hard negatives. Each color represents a malware family.}
      \label{fig:embedding_space_comparison_AndMal}
    \end{figure*}

        \begin{figure*}[!t]
      \centering
      \begin{subfigure}[b]{0.3\textwidth}
        \includegraphics[width=\textwidth]{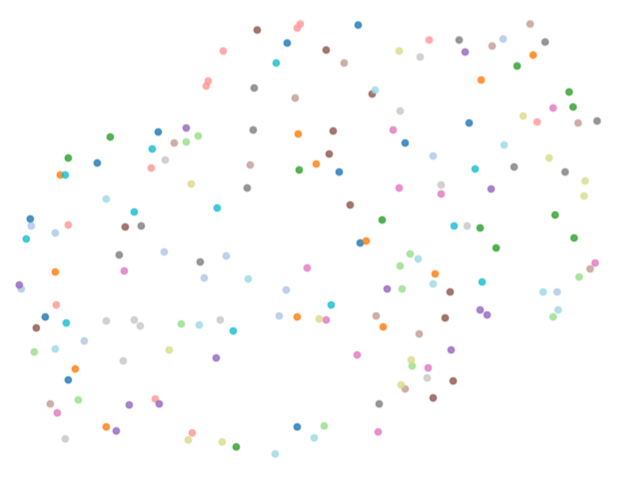}
        \caption{Pre-trained}
        \label{fig:pretrained}
      \end{subfigure}
      \hfill
      \begin{subfigure}[b]{0.3\textwidth}
        \includegraphics[width=\textwidth]{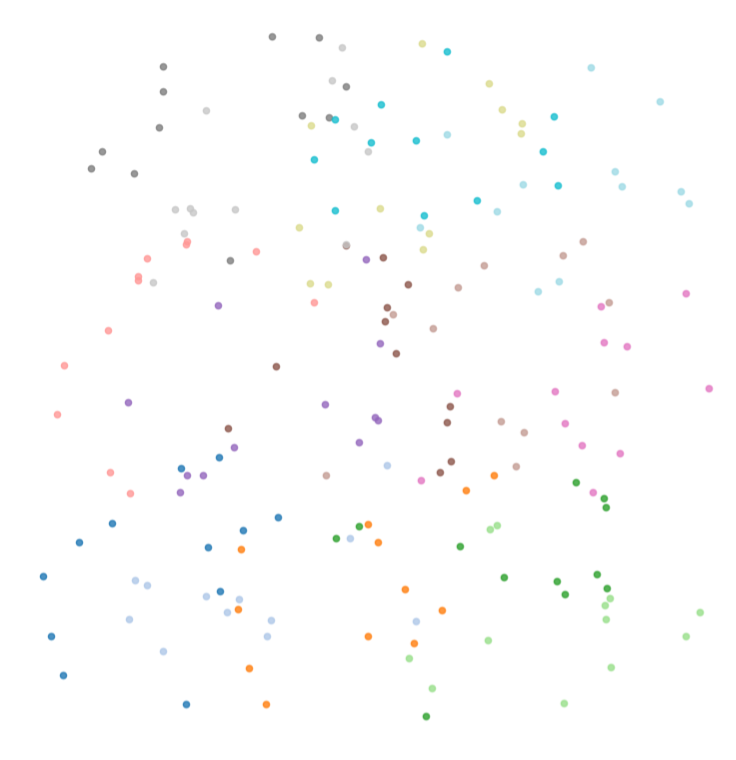}
        \caption{Random Negatives (CFT)}
        \label{fig:random_cft}
      \end{subfigure}
      \hfill
      \begin{subfigure}[b]{0.3\textwidth}
        \includegraphics[width=\textwidth]{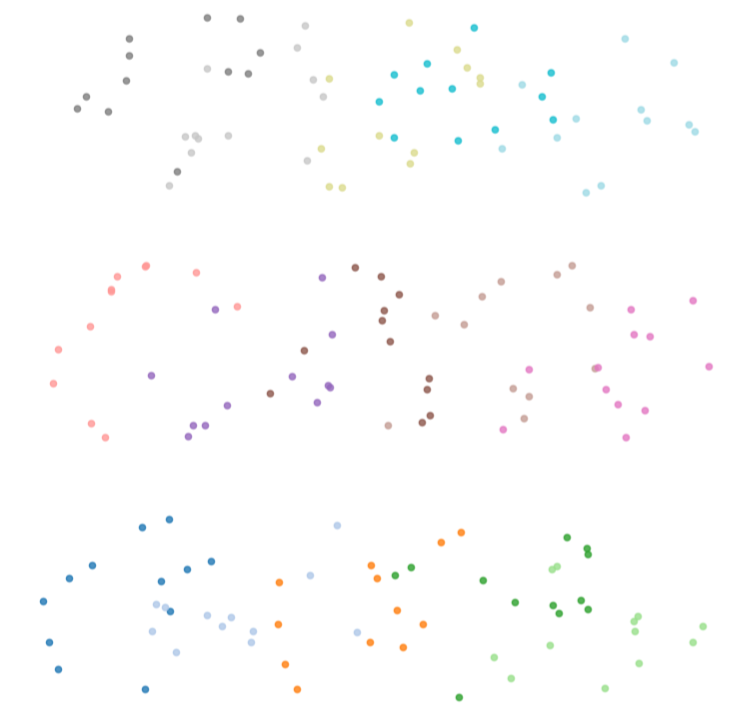}
        \caption{Similarity-based Negatives (CFT)}
        \label{fig:similarity_cft}
      \end{subfigure}
      \caption{\small Comparison of LLaMA-3.1-8B embedding spaces for BODMAS dataset: pre-trained vs. contrastive fine-tuning with random and similarity-based hard negatives. Each color represents a malware family.}
      \label{fig:embedding_space_comparison_BODMAS}
    \end{figure*}
    
    \subsection{Contrastive Fine-Tuning}
    
        Contrastive Fine-Tuning is conducted using the InfoNCE loss~\cite{oord2018representation} defined as:
    
        \begin{equation*}
            \mathcal{L}_{\text{InfoNCE}} = -\log \frac{\exp\left(\text{sim}(\mathbf{z}_i, \mathbf{z}_j)/\tau\right)}{\sum_{k=1}^{N} \exp\left(\text{sim}(\mathbf{z}_i, \mathbf{z}_k)/\tau\right)}
        \end{equation*}
        
        where $\mathbf{z}_i$ and $\mathbf{z}_j$ are anchor and positive embeddings, respectively, $\operatorname{sim}(\cdot,\cdot)$ denotes cosine similarity, $\tau$ is the temperature hyperparameter set to 0.07, and $N$ represents the total number of positive and negative samples per training batch. We use the AdamW optimizer with a learning rate of $1\times10^{-5}$, a batch size of 32, and train for 1 epoch across all LLM models and datasets.

    \subsection{Embedding Generation}
        After fine-tuning, embeddings are generated from the textual descriptions produced by each LLM. For the smaller LlaMA model, the embeddings are 2048-dimensional, the LlaMA-3.2-3B model contains 3072 embedding dimensions, while larger models produce embeddings of 4096 dimensions. Embeddings are derived by mean pooling the final hidden-state representations of the fine-tuned models.
    
    \subsection{Multimodal Classifier}
    
        Our multimodal classifier processes two distinct modalities: behavioral attributes (140-dimensional for CIC-AndMal-2020, and 64-dimensional for BODMAS) and embeddings (2048 or 4096-dimensional). The behavioral attributes (e.g., 64-dimensional for BODMAS) are projected into a 128-dimensional latent space using two fully connected layers with ReLU activation. Embeddings (2048–4096D) are reduced to 128D via a linear layer and then concatenated for a 256D-layer.
    
    \subsection{MAML Framework and Knowledge Distillation}
    
        We evaluate embedding effectiveness using Model-Agnostic Meta-Learning (MAML)~\cite{finn2017model} in a few-shot setting. Each malware family classification task is divided into support and query sets, containing 10 support and 20 query samples per family. In the inner loop, the multimodal classifier rapidly adapts to the support set by optimizing a cross-entropy loss. The outer loop aggregates gradients from query sets across tasks to update parameters, enabling rapid generalization to unseen malware samples.
    
        In addition, we incorporate a knowledge distillation step into MAML classifier training. A teacher model trained solely on dynamic attributes (binary features) provides soft labels to guide the multimodal (student) model. This distillation step encourages the multimodal model to leverage both embedding and dynamic attributes effectively, further enhancing classification accuracy.

\section{Results}\label{section:results}

    \subsection{Embedding Quality Evaluation}
    
        We first evaluate the quality of the embeddings produced by our contrastive fine-tuning method on both the \textit{CIC-AndMal-2020} and \textit{BODMAS} datasets. Figures~\ref{fig:embedding_space_comparison_AndMal} and~\ref{fig:embedding_space_comparison_BODMAS} show the embedding spaces for each dataset, comparing the pre-trained model with contrastive fine-tuning using random and similarity-based hard negatives.  While traditional CFT improves intra-family cohesion (clustering positives closer), it struggles to effectively separate hard negatives due to heuristic sampling (e.g., random or class-based negatives). Our similarity-based CFT, however, explicitly optimizes the embedding space to both cluster positives and push semantically proximate hard negatives apart, as evidenced by reduced overlap between families like Slocker and SmsSpy in Figure~\ref{fig:embedding_space_comparison_AndMal}. To quantify the practical utility of these embeddings, we employ RAGAS metrics (Answer Correctness and Similarity) to assess the quality of LLM-generated malware descriptions. As shown in Table~\ref{tab:rag_evaluation}, similarity-based CFT consistently outperforms other strategies. Notably, similarity-based CFT narrows the gap between correctness and similarity scores (e.g., Mistral-7B-v0.1 on BODMAS: 70.73 vs. 87.27), indicating that refined embeddings enable LLMs to generate descriptions that are both accurate and aligned with expert narratives. This alignment is critical for human analysts, as it ensures descriptions are not only machine-actionable but also interpretable. 

\begin{table*}[!t]
\renewcommand{\arraystretch}{1.3}
\caption{RAG Evaluation: Answer Correctness and Similarity Across Fine-Tuning Strategies. Measurement values in (\%)}
\label{tab:rag_evaluation}
\centering
\begin{tabular}{lcccccc}
\multicolumn{7}{c}{\textbf{ANDMAL}} \\
\hline
\bfseries Model & 
\multicolumn{2}{c}{\bfseries Pre-trained} & 
\multicolumn{2}{c}{\bfseries Random Negative CFT} & 
\multicolumn{2}{c}{\bfseries Similarity-Based CFT} \\
& \bfseries Correctness & \bfseries Similarity & \bfseries Correctness & \bfseries Similarity & \bfseries Correctness & \bfseries Similarity \\
\hline\hline
LLaMa-3.2-3B & 57.72\% & 78.11\% & 59.13\% & 77.99\% & \textbf{66.92\%} & \textbf{79.51\%} \\
LLaMa-3.1-8B & 56.72\% & 77.11\% & 59.28\% & 76.96\% & \textbf{69.79\%} & \textbf{80.57\%} \\
\hline
\multicolumn{7}{c}{\textbf{BODMAS}} \\
\hline\hline
LLaMa-3.2-1B & 61.88\% & 79.27\% & 64.88\% & 80.41\% & \textbf{66.88\%} & \textbf{83.57\%} \\
LLaMa-3.2-3B & 60.75\% & 83.66\% & 62.92\% & 84.95\% & \textbf{79.12\%} & \textbf{89.57\%} \\
LLaMa-3.1-8B & 59.46\% & 84.15\% & 61.59\% & 83.83\% & \textbf{73.81\%} & \textbf{88.41\%} \\
Mistral-7B-v0.1 & 58.28\% & 80.57\% & 50.15\% & 80.31\% & \textbf{70.73\%} & \textbf{87.27\%} \\
\hline
\end{tabular}
\end{table*}

\subsection{Malware Family Classification Results (MAML)}

We now evaluate the practical impact of our similarity-based contrastive embeddings on malware family classification tasks using the MAML framework. Tables~\ref{tab:maml_results_andmal} and~\ref{tab:maml_results_bodmas} summarize the classification accuracy obtained on CIC-AndMal-2020 and BODMAS datasets, respectively, comparing three embedding strategies: behavioral attribute-only baseline, pre-trained embeddings, and embeddings from our proposed contrastive fine-tuning approach.

\begin{table}[!t]
    \renewcommand{\arraystretch}{1.3}
    \caption{Malware Classification Accuracy on CIC-AndMal-2020 with MAML (\%)}
    \label{tab:maml_results_andmal}
    \centering
    \begin{tabular}{lc}
        \hline
        \bfseries Method & \bfseries Accuracy (\%) \\ 
        \hline\hline
        Behavioral Attributes (Baseline) & 42.00\% \\[4pt]
        \hline
        Pre-trained Embeddings & \\
        \hline
        \quad LLaMA-3.2-1B & 26.22\% \\
        \quad LLaMA-3.2-3B & 26.48\% \\
        \quad LLaMA-3.1-8B & 21.17\% \\
        \quad Mistral-7B-v0.1 & 26.23\% \\[4pt]
        \hline
        \textbf{Contrastive FT Embeddings (Ours)} & \\
        \hline
        \quad LLaMA-3.2-1B & \textbf{57.54\%} \\
        \quad LLaMA-3.2-3B & \textbf{58.59\%} \\
        \quad LLaMA-3.1-8B & \textbf{63.15\%} \\
        \quad Mistral-7B-v0.1 & \textbf{53.33\%} \\
        \hline
    \end{tabular}
\end{table}

\begin{table}[!t]
    \renewcommand{\arraystretch}{1.3}
    \caption{Malware Classification Accuracy on BODMAS with MAML (\%)}
    \label{tab:maml_results_bodmas}
    \centering
    \begin{tabular}{lc}
        \hline
        \bfseries Method & \bfseries Accuracy (\%) \\ 
        \hline\hline
        Behavioral Attributes (Baseline) & 32.00\% \\[4pt]
        \hline
        Pre-trained Embeddings & \\
        \hline
        \quad LLaMA-3.2-1B & 31.63\% \\
        \quad LLaMA-3.2-3B & 32.72\% \\
        \quad LLaMA-3.1-8B & 26.38\% \\
        \quad Mistral-7B-v0.1 & 33.87\% \\[4pt]
        \hline
        \textbf{Contrastive FT Embeddings (Ours)} & \\
        \hline
        \quad LLaMA-3.2-1B & \textbf{40.34\%} \\
        \quad LLaMA-3.2-3B & \textbf{48.27\%} \\
        \quad LLaMA-3.1-8B & \textbf{45.28\%} \\
        \quad Mistral-7B-v0.1 & \textbf{47.54\%} \\
        \hline
    \end{tabular}
\end{table}

The results demonstrate that embeddings generated through our proposed contrastive fine-tuning significantly outperform the baseline methods. On the CIC-AndMal-2020 dataset (Table~\ref{tab:maml_results_andmal}), our similarity-based fine-tuning achieves accuracy improvements ranging from approximately 11\% to over 20\% relative to the behavioral attribute-only baseline, and surpasses pre-trained embeddings by more than 25\% in all evaluated models. Notably, the larger LLaMA-3.1-8B model demonstrates the highest improvement, reaching 63.15\% accuracy, illustrating that richer embedding spaces greatly benefit from our contrastive optimization strategy.

Similarly, the BODMAS dataset (Table~\ref{tab:maml_results_bodmas}) exhibits notable improvements in classification accuracy. Here, our method achieves performance gains of approximately 8\% (LLaMA-3.2-1B) to 15\% (LLaMA-3.2-3B) over pre-trained embeddings, confirming consistent effectiveness across diverse datasets and indicating the robust generalization capabilities of the contrastively fine-tuned embeddings. Although baseline attribute-only accuracy on BODMAS is slightly lower than traditional BODMAS results due to the reduced-feature approach, integrating our embeddings consistently enhances model performance, underscoring the practical advantage of our method in multimodal malware classification.

Notably, raw embeddings from pre-trained LLMs consistently underperform even the simple behavioral attribute baseline. For instance, pre-trained LLaMA-3.1-8B achieves only 21.17\% accuracy on the CIC-AndMal-2020 dataset, significantly below the baseline of 42.00\%. This underscores the necessity of our CFT strategy, as raw embeddings alone fail to capture discriminative semantic features critical for cybersecurity relevance. Consequently, the substantial performance gains observed with our contrastively fine-tuned embeddings validate their effectiveness and highlight their practical utility in real-world malware classification scenarios.

    \subsection{Ablation Studies on Negative Selection}
    
        To validate our proposed negative sample selection strategy, we conducted ablation studies comparing two methods: random negative sampling and our similarity-based hard negative sampling approach. Table~\ref{tab:ablation_negatives} summarizes the classification accuracy obtained using both methods across all four LLM models and both datasets.

        \begin{table}[!t]
            \renewcommand{\arraystretch}{1.3}
            \caption{Ablation Study: Random vs. Similarity-Based Negatives on MAML Accuracy (\%)}
            \label{tab:ablation_negatives}
            \centering
            \begin{tabular}{lcc}
            \hline
                \bfseries Method & \bfseries Random Negatives (\%) & \bfseries Similarity-Based \\ & & \textbf{Negatives} (\%) \\
                \hline\hline
                \multicolumn{3}{c}{\textbf{CIC-AndMal-2020}} \\
                \hline
                LLaMA-3.2-1B & 40.48\% & \textbf{57.54\%} \\
                LLaMA-3.2-3B & 43.91\% & \textbf{58.59\%} \\
                LLaMA-3.1-8B & 40.34\% & \textbf{63.15\%} \\
                Mistral-7B-v0.1 & 42.75\% & \textbf{53.33\%} \\
                \hline
                \multicolumn{3}{c}{\textbf{BODMAS}} \\
                \hline
                LLaMA-3.2-1B & 36.26\% & \textbf{40.34\%} \\
                LLaMA-3.2-3B & 36.43\% & \textbf{48.27\%} \\
                LLaMA-3.1-8B & 39.6\% & \textbf{45.28\%} \\
                Mistral-7B-v0.1 & 29.24\% & \textbf{47.54\%} \\
                \hline

            \end{tabular}
            
            
        \end{table}
        As indicated in Table~\ref{tab:ablation_negatives}, our similarity-based negative sampling consistently outperforms random negative sampling across all models and datasets. On the CIC-AndMal-2020 dataset, our approach achieves performance gains ranging from approximately 10\% (Mistral-7B-v0.1) to over 20\% (LLaMA-3.1-8B), demonstrating substantial improvement in distinguishing closely related malware families. This suggests that similarity-based negative sampling is particularly beneficial for scenarios characterized by significant semantic overlap, as it effectively trains the model to identify and differentiate subtle semantic variations.Larger models (e.g., LLaMA-3.1-8B) achieve greater gains with similarity-based negatives, suggesting that capacity is key to exploiting fine-grained semantic differences. Similarly, results from the BODMAS dataset further confirm the robustness and efficacy of our proposed method. While the overall accuracy gains are slightly more moderate compared to the CIC-AndMal-2020 dataset, the improvements remain significant, ranging from roughly 4\% (LLaMA-3.2-1B) up to approximately 18\% (Mistral-7B-v0.1). The consistent superiority across diverse models and both datasets highlights the generalizability of our similarity-based negative sampling approach.

\section{Conclusion}\label{section:conclusion}
    This paper introduces a novel contrastive fine-tuning framework for malware family classification, which refines textual embeddings by strategically selecting hard negative samples based on cosine similarity. By fusing contrastive learning with multimodal meta-learning, our method optimizes LLMs to generate attribute-aware descriptions while aligning them with structured behavioral features. Experiments demonstrate that our approach outperforms baseline methods and regular CFT across CIC-AndMal-2020 and BODMAS datasets, enabling robust generalization to unseen malware variants. Moreover, our evaluation with RAGAS metrics confirms that similarity-based negative sampling significantly improves the quality of human-readable malware descriptions. Specifically, refined embeddings produced by our approach lead to descriptions that achieve higher correctness scores and better alignment with expert-generated narratives, enhancing both their accuracy and interpretability. Such improvements in human readability and interpretability are critical, ensuring generated descriptions are valuable not only for automated systems but also for cybersecurity analysts.
        
\section*{Acknowledgment}
    We would like to thank flaticon.com for icons in Fig. \ref{fig:tldr}.

\bibliographystyle{unsrt}
\bibliography{bibliography}

\end{document}